\documentclass{PoS}

\renewcommand{\logo}{\raisebox{-5mm}{\hbox{}}}

\makeatletter
\global\setbox\PoSpaper@url\hbox{}
\makeatother

\title{Recent results from the \textsc{Majorana Demonstrator}}

\ShortTitle{Recent results from the \textsc{Majorana Demonstrator}}
\author{\speaker{J.~Myslik},$^a$
S.I.~Alvis,$^{b}$
I.J.~Arnquist,$^{c}$
F.T.~Avignone~III,$^{de}$
A.S.~Barabash,$^{f}$
C.J.~Barton,$^{g}$
F.E.~Bertrand,$^{e}$
T.~Bode,$^{h}$
B.~Bos,$^{i}$ 
V.~Brudanin,$^{j}$ 
M.~Busch,$^{kl}$ 
M.~Buuck,$^{b}$  
T.S.~Caldwell,$^{ml}$ 
Y-D.~Chan,$^{a}$ 
C.D.~Christofferson,$^{i}$ 
P.-H.~Chu,$^{n}$  
C. Cuesta,$^{b}$\footnote{Present address: Centro de Investigaciones Energ\'{e}ticas, Medioambientales y Tecnol\'{o}gicas, CIEMAT 28040, Madrid, Spain}
J.A.~Detwiler,$^{b}$ 
C.~Dunagan,$^{i}$ 
Yu.~Efremenko,$^{oe}$ 
H.~Ejiri,$^{p}$ 
S.R.~Elliott,$^{n}$ 
T.~Gilliss,$^{ml}$  
G.K.~Giovanetti,$^{q}$  
M.P.~Green,$^{rle}$  
J.~Gruszko,$^{s}$  
I.S.~Guinn,$^{b}$ 
V.E.~Guiseppe,$^{d}$
C.R.~Haufe,$^{ml}$ 
R.J.~Hegedus,$^{ml}$  
L.~Hehn,$^{a}$ 	
R.~Henning,$^{ml}$ 
D.~Hervas~Aguilar,$^{ml}$ 
E.W.~Hoppe,$^{c}$ 
M.A.~Howe,$^{ml}$ 
K.J.~Keeter,$^{t}$ 
M.F.~Kidd,$^{u}$ 	
S.I.~Konovalov,$^{f}$ 
R.T.~Kouzes,$^{c}$ 
A.M.~Lopez,$^{o}$ 	
R.D.~Martin,$^{v}$ 	
R.~Massarczyk,$^{n}$
S.J.~Meijer,$^{ml}$ 	
S.~Mertens,$^{hw}$ 
G.~Othman,$^{ml}$  
W.~Pettus,$^{b}$ 
A.~Piliounis,$^{v}$  
A.W.P.~Poon,$^{a}$ 
D.C.~Radford,$^{e}$ 
J.~Rager,$^{ml}$ 	
A.L.~Reine,$^{ml}$ 	
K.~Rielage,$^{n}$ 
N.W.~Ruof,$^{b}$
B.~Shanks,$^{e}$ 
M.~Shirchenko,$^{j}$ 
D.~Tedeschi,$^{d}$ 
R.L.~Varner,$^{e}$ 
S.~Vasilyev,$^{j}$ 	
Vasundhara,$^{v}$  
B.R.~White,$^{n}$ 	
J.F.~Wilkerson,$^{mle}$     
C.~Wiseman,$^{b}$ 		
W.~Xu,$^{g}$  
E.~Yakushev,$^{j}$ 
C.-H.~Yu,$^{e}$ 
V.~Yumatov,$^{f}$ 
I.~Zhitnikov,$^{j}$ 
 and B.X.~Zhu$^{n}$
\\
\llap{$^{a}$} Nuclear Science Division, Lawrence Berkeley National Laboratory, Berkeley, CA, USA\\
\llap{$^{b}$} Center for Experimental Nuclear Physics and Astrophysics, 
and Department of Physics, University of Washington, Seattle, WA, USA\\
\llap{$^{c}$} Pacific Northwest National Laboratory, Richland, WA, USA\\
\llap{$^{d}$} Department of Physics and Astronomy, University of South
Carolina, Columbia, SC, USA \\
\llap{$^{e}$} Oak Ridge National Laboratory, Oak Ridge, TN, USA\\
\llap{$^{f}$} National Research Center ``Kurchatov Institute'' Institute for
Theoretical and Experimental Physics, Moscow, Russia \\
\llap{$^{g}$} Department of Physics, University of South Dakota, Vermillion,
SD, USA\\
\llap{$^{h}$} Max-Planck-Institut f\"{u}r Physik, M\"{u}nchen, Germany\\
\llap{$^{i}$} South Dakota School of Mines and Technology, Rapid City, SD,
USA\\
\llap{$^{j}$} Joint Institute for Nuclear Research, Dubna, Russia\\
\llap{$^{k}$} Department of Physics, Duke University, Durham, NC, USA\\
\llap{$^{l}$} Triangle Universities Nuclear Laboratory, Durham, NC, USA\\
\llap{$^{m}$} Department of Physics and Astronomy, University of North
Carolina, Chapel Hill, NC, USA\\
\llap{$^{n}$} Los Alamos National Laboratory, Los Alamos, NM, USA\\
\llap{$^{o}$} Department of Physics and Astronomy, University of Tennessee,
Knoxville, TN, USA\\
\llap{$^{p}$} Research Center for Nuclear Physics, Osaka University, Ibaraki,
Osaka, Japan\\
\llap{$^{q}$} Department of Physics, Princeton University, Princeton, NJ, USA\\
\llap{$^{r}$} Department of Physics, North Carolina State University, Raleigh,
NC, USA\\
\llap{$^{s}$} Department of Physics, Massachusetts Institute of Technology,
Cambridge, MA, USA\\
\llap{$^{t}$} Department of Physics, Black Hills State University, Spearfish,
SD, USA\\
\llap{$^{u}$} Tennessee Tech University, Cookeville, TN, USA\\
\llap{$^{v}$} Department of Physics, Engineering Physics and Astronomy, Queen's
University, Kingston, ON, Canada \\
\llap{$^{w}$} Physik Department, Technische Universit\"{a}t, M\"{u}nchen,
Germany\\
E-mail: \email{jwmyslik@lbl.gov}}
\abstract{The \textsc{Majorana Demonstrator} is an experiment constructed to search for
neutrinoless double-beta decay in $^{76}$Ge and to demonstrate the feasibility to
deploy a large-scale experiment in a phased and modular fashion. It consists of
two modules of natural and $^{76}$Ge-enriched germanium detectors totalling
44.1 kg, operating at the 4850' level of the Sanford Underground Research
Facility in Lead, South Dakota, USA. Commissioning of the experiment
began in June 2015, followed by data production with the full detector
array in August 2016. The ultra-low background and record energy resolution achieved by the
\textsc{Majorana \mbox{Demonstrator}}  enable a sensitive neutrinoless double-beta decay search, as well
as additional searches for physics beyond the Standard Model. I will discuss
the design elements that enable these searches, along with the latest results,
focusing on the neutrinoless double-beta decay search. I will also discuss the
current status and the future plans of the \textsc{Majorana Demonstrator}, as well as
the plans for a future tonne-scale $^{76}$Ge experiment.}

\dedicated{This material is
based upon work supported by the U.S. Department of Energy, Office of Science,
Office of Nuclear Physics, the Particle Astrophysics and Nuclear Physics
Programs of the National Science Foundation, the Russian Foundation for Basic
Research, the Natural Sciences and Engineering Research Council of Canada, the
Canada Foundation for Innovation John R. Evans Leaders Fund, the National
Energy Research Scientific Computing Center, and the Oak Ridge Leadership
Computing Facility, and the Sanford Underground
Research Facility.}

\FullConference{The 39th International Conference on High Energy Physics (ICHEP2018)\\
		4-11 July, 2018\\
		Seoul, Korea}

\begin{document}

\section{Introduction}
The \textsc{Majorana Demonstrator} \cite{Abgrall:2013rze} is a $^{76}$Ge neutrinoless double-beta
decay ($0\nu\beta\beta$-decay) experiment, currently operating at the 4850' level of the Sanford
Underground Research Facility (SURF).  It consists of 44.1~kg of p-type point
contact germanium crystal detectors, 29.7~kg enriched to $88.1\% \pm
0.7\%$
$^{76}$Ge.  They are installed in two vacuum cryostats, shielded with copper,
lead, and polyethylene, with an active muon veto.  Ultra-clean materials were
used in its construction, including electroformed copper. Care was
taken to limit the cosmogenic exposure of the germanium following enrichment,
fabricated detectors, and any electroformed copper components, which required
above-ground fabrication processes. The resulting low backgrounds, coupled with low noise electronics, 
enable sensitive searches for other physics beyond the Standard Model \cite{Abgrall:2016tnn,Alvis:2018yte}.

\section{Pulse shape based background rejection techniques}
In addition to shielding and ultra-clean material selection, analysis of
detector pulses also provides a means to reject backgrounds.  Two features of
the p-type point contact detectors \cite{Abgrall:2013rze} are their range of drift times and localized 
weighting potential.  Background $\gamma$-ray events often deposit energy at multiple points 
in the detector, while double-beta decays release
their energy at a single site (within $\sim1$~mm of the event location).  Through examination of the current
pulse amplitude versus the total energy, multi-site events can be distinguished
from single-site events and rejected.

Another technique concerns an observed energy-degraded $^{210}$Po alpha background that
studies indicate likely originates from $^{210}$Pb near the
detector's point contact.  This alpha must pass through the passivated surface
of a detector, and the slow drift of charges along the passivated surface produces a
distinctive waveform, allowing this background to be identified and rejected.

\section{Neutrinoless double-beta decay results}
The \textsc{Majorana Demonstrator} has been operating in a stable
configuration since the spring of 2017.  Changes between earlier configurations
require the data to be split into multiple datasets (DS).  In particular, there is
higher background in DS0, taken prior to the installation of the
inner copper shield; and in DS5a,
when a single DAQ system was used and the pulse shape analysis was degraded by high
electronic noise until the shielding and grounding path were completed.
Therefore, DS0 and DS5a are excluded from the lowest background configuration,
which better represents achieved background rates. Simulations and assays indicate that when prominent background $\gamma$ lines
are excluded, the spectrum from 1950~keV to 2350~keV is flat in shape.
Therefore, to estimate the background, a $\pm5$~keV window is excluded around each of
these lines ($^{208}$Tl: 2104~keV; $^{214}$Bi: 2118~keV, and 2204~keV), as well as $Q_{\beta\beta} =
2039$~keV (the energy resolution achieved at Q$_{\beta\beta}$ is 2.5~keV FWHM).  
The remaining $360$-keV window is used to estimate the
background rate at $Q_{\beta\beta}$. To determine a limit on the
$0\nu\beta\beta$-decay half life, the
360-keV window plus the $Q_{\beta\beta}$ region is analyzed with a nominal
profile likelihood method, searching on top of a flat background for a peak
at $Q_{\beta\beta}$ modelled by an exponentially modified Gaussian
distribution.  Independent analyses are also peformed to cross-check the
validity of the statistical result, using the Feldman-Cousins method, the CLs method (a
modified profile likelihood method) and a Bayesian Markov-Chain Monte Carlo
(with a flat prior assigned to the $0\nu\beta\beta$-decay rate).

In 2017 \cite{Aalseth:2017btx}, open data in DS0-5b (9.95~kg$\cdot$yr of enriched germanium exposure)
was analyzed, finding a
background rate of $17.8\pm3.6$~counts/(FWHM$\cdot$t$\cdot$yr), and
$T_{1/2}^{0\nu} > 1.9 \times 10^{25}$~yr (median sensitivity:  $2.1
\times 10^{25}$~yr) at the 90\% confidence level.  There was 5.24~kg$\cdot$yr
of exposure in the lowest background configuration, which had a background rate
of $4.0^{+3.1}_{-2.5}$~counts/(FWHM$\cdot$t$\cdot$yr) or 
$1.6^{+1.2}_{-1.0}\times10^{-3}$~counts/(keV$\cdot$kg$\cdot$yr).

The \textsc{Majorana Demonstrator} employed a statistical blindness scheme,
whereby 31 hours of open data is followed by 93 hours of blind data with the
full spectrum blinded. Unblinding was performed in three phases to enable data
quality and consistency checks.  First, the region outside the background
evaluation window was opened in order to finalize the run and channel
selection.   Then, the background evaluation window was opened to measure the
background.  Finally, the Q$_{\beta\beta}$ region was opened, and the half-life
limit is set.  Multi-detector events and the spectrum below 100~keV remain blind
for other studies.

For the 2018 analysis, all open and blind data taken before April 15, 2018 was
analyzed.  This added 16.1~kg$\cdot$yr of exposure, for a total of
26~kg$\cdot$yr of enriched germanium exposure. Analysis of this exposure found
a background rate of  $15.4\pm2.0$~counts/(FWHM$\cdot$t$\cdot$yr) and a
half-life limit of $T_{1/2}^{0\nu} > 2.7 \times 10^{25}$~yr (median
sensitivity:  $4.8 \times 10^{25}$~yr) at the 90\% confidence level.  The
lowest background configuration had an enriched germanium exposure of
21.3~kg$\cdot$yr, and a background rate of
$11.9\pm2.0$~counts/(FWHM$\cdot$t$\cdot$yr) or 
$(4.7\pm0.8)\times10^{-3}$~counts/(keV$\cdot$kg$\cdot$yr).

\section{Conclusions}
The \textsc{Majorana Demonstrator} has completed construction, and has been
taking data in its final configuration since the spring of 2017.  With a
combined (open+blind) exposure of 26~kg$\cdot$yr, the \textsc{Majorana
Demonstrator} places a lower limit on the $0\nu\beta\beta$-decay
half-life of $2.7\times 10^{25}$~yr (median sensitivity: $4.8\times 10^{25}$~yr) at
the 90\% confidence level.  Ultimately, the \textsc{\mbox{Majorana} Demonstrator}
expects to reach an exposure of 50-70~kg$\cdot$yr, pushing its sensitivity to a
half-life $\sim10^{26}$~yr.  With its goal of lower backgrounds, sensitivity beyond $10^{28}$ years will be
possible with the next-generation tonne-scale experiment, LEGEND.  Its first
phase, the 200-kg LEGEND-200, will begin taking data in 2021, and will have
sensitivity beyond $10^{27}$ years.

\bibliographystyle{JHEP}
\bibliography{ICHEP2018_Proceedings_MJD_Myslik}

%

\end{document}